\documentclass[prb,preprintnumbers,twocolumn,amsmath,amssymb,superscriptaddress]{revtex4}
\usepackage{graphicx}
\usepackage{color}
\usepackage{dcolumn}
\usepackage{bm}
\usepackage{psfrag}

\begin{document}
\title{On the application of Mattis-Bardeen theory in strongly disordered superconductors} 
\author{G.~Seibold} 
\affiliation{Institut F\"ur Physik, BTU Cottbus, PBox 101344, 03013 Cottbus,
Germany}
\author{L.~Benfatto} 
\affiliation{ISC-CNR and Department of Physics, University of Rome ``La
  Sapienza'',\\ Piazzale Aldo Moro 5, 00185, Rome, Italy}
\author{C.~Castellani}
\affiliation{ISC-CNR and Department of Physics, University of Rome ``La
  Sapienza'',\\ Piazzale Aldo Moro 5, 00185, Rome, Italy}
\date{\today}

\begin{abstract}
The low energy optical conductivity of conventional superconductors is usually
well described by Mattis-Bardeen (MB) theory which predicts the onset of absorption
above an energy corresponding to twice the superconducing (SC) gap parameter $\Delta$. 
Recent experiments 
on strongly disordered superconductors have challenged the application of the MB formulas
due to the occurrence of additional spectral weight at low energies below $2\Delta$. 
Here we identify three crucial items which have to be included in the analysis of
optical-conductivity data for these systems: (a) the correct identification of the 
optical threshold in the Mattis-Bardeen theory, and its relation with the gap value 
extracted from the measured density of states, (b) the gauge-invariant evaluation of 
the  current-current response function, needed to account for the optical absorption 
by SC collective modes, and (c) the inclusion into the MB formula of the energy 
dependence of the density of states present already above $T_c$. By computing the 
optical conductvity in the disordered attractive Hubbard model we analyze the relevance 
of all these items, and we provide a compelling scheme for the analysis and interpretation 
of the optical data in real materials. 
\end{abstract}


\maketitle

\section{Introduction}
The Bardeen-Cooper-Schrieffer (BCS) theory of superconductivity \cite{bcs} is probably one 
of the most successful examples in condensed matter of a microscopic approach able to 
describe a phase transition via the modification of the quasiparticle spectrum. In particular, the opening of a
superconducting (SC) gap $\Delta$ below $T_c$, along with the emergence of a purely 
diamagnetic response (Meissner effect), are the basic ingredients required to interpret
 the thermodynamic and transport properties of conventional superconductors. 
 However, to understand the finite-frequency optical response in the superconducting 
 state, additional effects due to the presence of disorder must be included as well. 
 The straightforward extension of BCS theory in the presence of non-magnetic impurities
was indeed developed soon after by Mattis and Bardeen (MB) \cite{mb} and independently
 by Abrikosov and Gorkov \cite{ag58}. In this case the single-particle excitations are 
 modified not only by the pairing, but also by the broadening $\Gamma$ of the energy 
 levels due to scattering by impurities which is present already above $T_c$. 
 The consequences for the optical absorption below $T_c$, i.e. the real part of the 
 optical conductivity $\sigma_1(\omega)$, have been worked out explicitly in the 
 milestone MB work \cite{mb}. While in the clean case all carriers (density $n$) contribute
 to the superfluid delta-like response at zero frequency,
in the dirty limit only a fraction $n_s\sim n (\Delta/\Gamma)$ of the carriers condenses
in the superfluid state. By sum-rule conservation the reduction of $n_s$ has a direct 
counterpart in the emergence of a finite-frequency optical absorption. Indeed, the MB model
predicts the onset of absorption at $T=0$ above the superconducting gap $2\Delta$, while
for finite temperatures $T<T_c$ an additional quasiparticle contribution appears also
below $2\Delta$.
In the intermediate disorder regime  $\Gamma\sim\Delta$ the analytical MB formula does 
not apply, but a finite-frequency absorption above $2\Delta$  still survives\cite{zimmermann91}.

\begin{figure}[htb]
\includegraphics[width=8cm,clip=true]{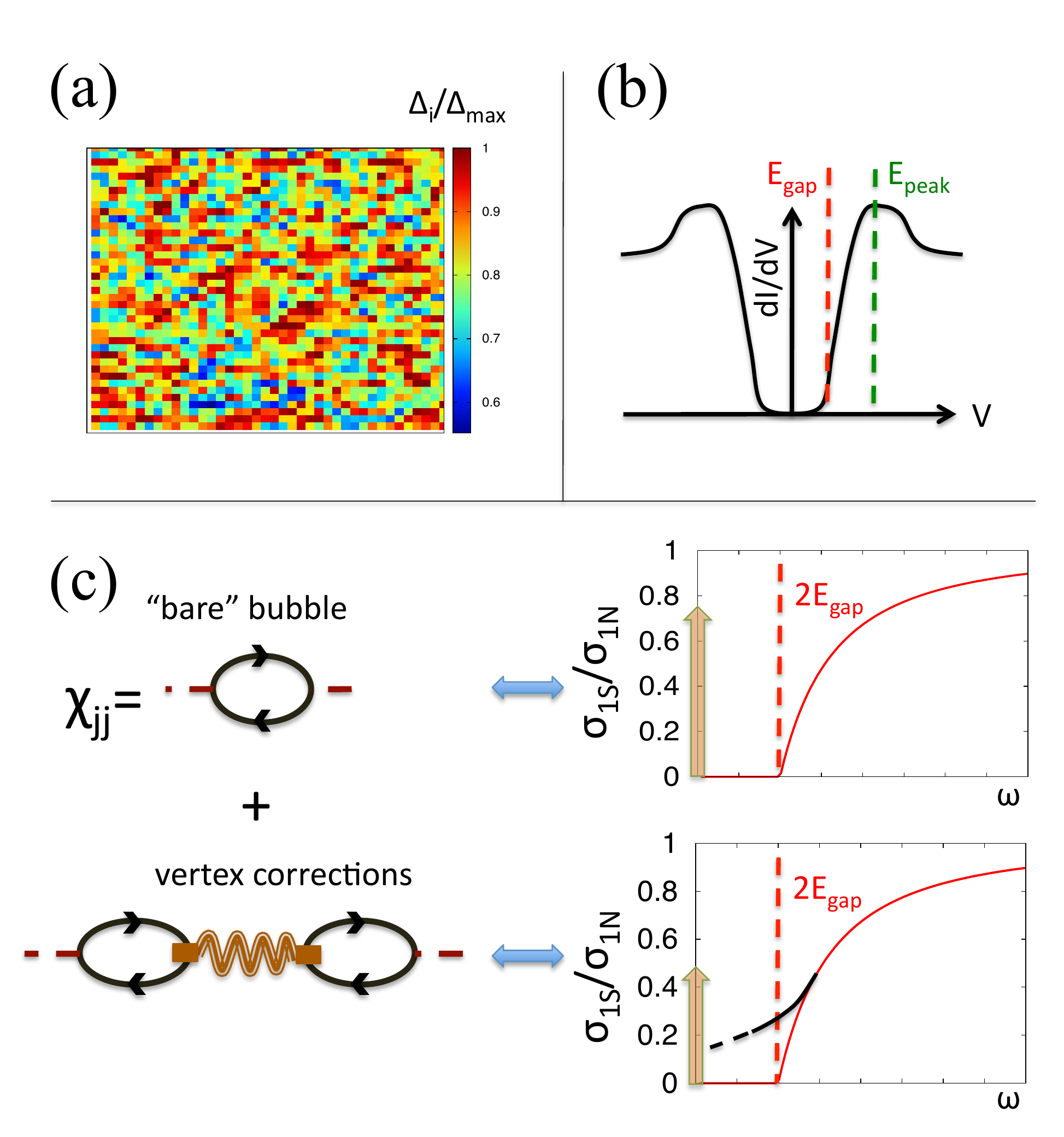}
\caption{Summary of the main spectral properties of strongly disordered superconductors. (a) Disorder induces a strong variation of the local SC order parameter $\Delta_i$, that is captured already at the level of the Bogoliubov-de-Gennes solutions for the disordered attractive Hubbard model\cite{trivedi_prb01,nandini_natphys11,seibold_prl12,lemarie_prb13}. Experimentally the variations occur as fluctuations of the coherence-peak height in the local DOS probed by STS\cite{sacepe11,lemarie_prb13,kam13}. (b) Sketch of the typical average tunneling DOS, as found numerically (see Fig.\ \ref{fig1} below)  and experimentally\cite{sacepe08,sacepe10,mondal10,sacepe11,chand12,kam13,roditchev13,samuely16}. The typical SC coherence peaks are strongly suppressed in intensity, and they are located at an energy scale $E_{peak}$ somehow larger than the minimun excitation energy $E_{gap}$. (c) Optical conductivity of disordered SC. Here the solid lines denote the SC Nambu Green function, the dashed lines the e.m. field and the wavy line the collective modes computed at RPA level, which build up the vertex corrections. The BCS approximation is equivalent to compute  the 
"bare bubble" diagram, which accounts only for the breaking of Cooper pairs by the incoming e.m. field. Even when the SC order parameter is inhomogeneous, as in panel (a), it leads to an optical conductivity with a hard threshold at $2E_{gap}$, that reproduces for intermediate disorder the usual Mattis-Bardeen results. The RPA vertex corrections account for the exchange of collective excitations between the Cooper pairs broken apart by the e.m. field. When these are included the optical conductivity acquires an extra contribution which piles up below $2E_{gap}$, stealing spectral weight from the superfluid peak at zero frequency, denoted here by an arrow. The black dashed line indicates that the precise form of $\sigma_{1s}$ as $\omega\rightarrow 0$ depends on the disorder level.  }
\label{figdiagrams}
\end{figure}

The MB theory successfully explains the microwave data in moderately disordered superconductors, as early experiments in  
indium and tin films demonstrated\cite{rug67,parks}. Later on it has been proven that the MB scheme can be 
extended to include also strong coupling effects\cite{nam67}, as observed e.g. in 
lead\cite{palmer68}, or a temperature-dependent scattering rate  of the residual (normal) 
quasiparticle excitations below $T_c$, as noticed in the analysis of Al films\cite{dressel_prb08}. 
 Only recently, studies on strongly disordered films of conventional 
superconductors\cite{armitage_prb07,driessen12,driessen13,frydman_natphys15,samuely15,armitage15},
granular superconductors\cite{bachar_jltp14,practh_prb16,scheffler16}, and even 
cuprate superconductors\cite{corson} have revealed systematic deviations from the MB 
paradigm in the form of an extra subgap absorption, that fairly exceeds the quasiparticle 
contribution of the MB theory. In the case of homogeneously disordered 
films\cite{armitage_prb07,driessen12,driessen13,frydman_natphys15,samuely15,armitage15} 
the identification of this subgap contribution also relies on the simultaneous estimate 
of the pairing energy scale from the tunnelling spectra, which by itself is a non trivial 
issue. Indeed, tunnelling measurements  performed by several 
groups\cite{sacepe08,sacepe10,mondal10,sacepe11,chand12,kam13,roditchev13,samuely16} revealed that in 
strongly-disordered films of NbN, InO$_x$ and TiN, where disorder induces a direct 
superconductor-to-insulator transition (SIT), the density of states  (DOS) shows 
significant deviations from the usual BCS form, as summarized in Fig.\ \ref{figdiagrams}a,b.  More specifically,  while BCS theory 
at $T\ll T_c$ predicts a square-root divergence of the DOS at  $\pm \Delta$, and the total 
suppression of the available states below the gap, the measurements reveal strongly suppressed
coherence peaks at an energy $E_{peak}$ and a finite DOS below it, with tails extending up to a somehow lower energy $E_{gap}$. In addition, the local density of states probed by 
means of scanning tunneling spectroscopy (STS) shows that these features vary on the
nanometer scale, leading to an inhomogeneous pattern of the local SC properties even in 
the presence of structurally homogeneous disorder\cite{sacepe10,kam13,roditchev13}. This 
emergent granularity bears also non-trivial space 
correlations\cite{sacepe10,kam13,lemarie_prb13}, with glassy features and quantum-percolation 
effects that have been interpreted theoretically as an outcome of the competition between
pair hopping and localization at the verge of the 
SIT\cite{trivedi_prb01,dubi_nat07,ioffe,nandini_natphys11,seibold_prl12,lemarie_prb13,mandal_prl13,seibold_prb15}. 
In this scenario, as the SIT is approached the pairing scale $E_{gap}$ stays finite even
when the average SC order parameter $\Delta\sim T_c$ softens, in agreement with tunneling 
observation\cite{sacepe10,mondal10,kam13}.   All these features, which cannot be captured 
with simple phenomenological models, as e.g. the Dynes formula \cite{dynes}, raise the non-trivial issue of the correct identification from the DOS of the energy scale to be compared with the measured  optical absorption threshold.

From the theoretical point of view, the failure of the MB theory at strong disorder is not
unexpected, since the MB theory describes only the effect of disorder on the quasiparticle
response, but it neglects the contribution of collective modes. 
This distinction relies on the usual diagrammatic expansion of the current-current correlation function $\chi_{jj}$, which enters the 
electromagnetic (e.m.) kernel   within the Kubo formalism (see Sec. II below for further details). 
At leading order $\chi_{jj}$ is given by a bubble diagram which represents the excitation of a particle-hole pair 
by the incoming e.m. field, see Fig.\ \ref{figdiagrams}c. At BCS level one just computes 
this "bare bubble" diagram by taking into account the modification of the electronic 
single-particle spectrum and the presence of anomalous electronic averages in the SC state, 
leading to all the well-known predictions of the BCS theory for the transport 
properties.\cite{schrieffer} On the other hand, below  $T_c$ also collective SC excitations are possible, 
connected to the amplitude and phase fluctuations of 
the SC order parameter around the BCS mean-field equilibrium value. The former one is a 
massive mode, also refered to as Higgs mode due to the analogy with the particle of the 
Standard Model\cite{weinberg}, while the second one is massless at long wavelength, since 
it represents the Goldstone mode of the $U(1)$ gauge symmetry breaking in the SC state\cite{nagaosa}. 
The effect of collective modes can be included within a gauge-invariant (GI) random-phase-approximation (RPA) by computing 
the vertex corrections to the current-current correlation function, that corresponds to 
account for all the intermediate processes between the particle-hole pairs excited by the 
e.m field before than they recombine, see Fig.\ \ref{figdiagrams}c. In the clean case and for transverse fields both the Higgs and phase
mode are decoupled from the current at long wavelength, even though the phase modes are crucial to restore the full gauge invariance of the longitudinal 
response function \cite{schrieffer,depalo_prb99,randeria_prb00,benfatto_prb04}. Thus, as long as one focuses on the transverse
e.m. response the BCS approximation successfully explains 
the physics of conventional clean superconductors, even though it is not gauge invariant. 

In the presence of disorder the computation of the SC e.m. response is more
complicate. First, disorder affects the single-particle excitations. This effect is
already present at the level of the bare-bubble approximation, and it is described by the
MB theory with some additional approximations, as e.g. a constant DOS of the electrons
in the normal state and a clean BCS-like DOS in the SC state. As we mentioned
before, this is not  always the case in disordered films, where the DOS at the Fermi
level is usually suppressed already above $T_c$\cite{sacepe10,mondal10,armitage15} and the BCS coherence peaks 
at $E_{peak}$ are smeared out in the SC state, with tails below it\cite{sacepe08,sacepe10,mondal10,kam13,roditchev13,samuely16}.  
Recently, it has been argued\cite{armitage15} that the latter effect can also be responsible for the
extra low-energy absorption in optics, once the MB theory is modified with a DOS adopted
from the Larkin-Ovchinnikov model\cite{larkin71}. In this view the MB response is argued to have still an optical 
threshold at $2E_{peak}$, with absorption tails below it due to the existence of in-gap states down to $2E_{gap}$. 
However, this analysis does not provide a good fit of the data, since it appears to underestimate 
the absorption in a large frequency range starting right above the optical threshold. 

In addition to the single-particle
effects, disorder affects also the optical observability of the collective modes, making them 
possible candidates to explain the experimental anomalies. So far, two main proposals have been put forward. From one side, it has been emphasized\cite{cea_prb14} that disorder mixes the response at finite and zero momentum, making the contribution of the RPA vertex corrections finite even for the physical transverse response functions. While it has been clearly proven\cite{cea_prb14} that collective modes can then lead to additional 
absorption with respect to the BCS result, it has not been clarified yet where this extra dissipation occurs with respect to the typical energies scale $E_{gap}, E_{peak}$ identified by STM.  A second proposal, put recently 
forward by Refs.\ [\onlinecite{podolsky_prb11,frydman_natphys15}], focuses instead on the possibility that the amplitude Higgs mode contributes to the extra absorption via optical processes beyond RPA level, present in principle already for the clean system\cite{podolsky_prb11,trivedi_prx14}. In this case the crucial role of disorder should be to change the nature itself of the mode, moving the Higgs resonance below the quasiparticle continuum at $2E_{gap}$,  restoring thus its relativistic dynamics\cite{varma_review,cea_prl15}. 
However, this interpretation has been recently questioned in Ref. \onlinecite{cea_prl15}, where it has been shown that a
strong overdamping of the Higgs mode persists even at strong disorder.

%

In this paper we present a detailed investigation of the optical properties
of disordered superconductors with the aim to clarify all the possible mechanisms leading
to significant deviations from the MB paradigm, and to provide a general scheme to interpret current experiments. We model the system by means of the
attractive Hubbard model with on-site disorder, that has been already
shown \cite{trivedi_prb01,dubi_nat07,nandini_natphys11,seibold_prl12,lemarie_prb13,mandal_prl13,seibold_prb15,cea_prl15} 
to reproduce many of the unconventional features observed experimentally in strongly
disordered thin films. By solving the  Bogoliubov-de Gennes (BdG) equations we can
describe the mean-field ground state in the presence of disorder. The 
optical conductivity is then computed both at the BCS level, i.e. as the bare-bubble
response, and by adding the RPA vertex corrections.  
The computation of the BCS response allows us to investigate how the inhomogeneity of the
SC properties triggered by strong disorder modifies the mean-field response with respect
to the predictions of the MB theory, and to identify the relevant energy scales for the optical absorption. 
In particular, we investigate the role of two separate effects observed in the quasiparticle DOS: (i) the smearing of the SC peak
due to inhomogeneity of the local SC order parameter $\Delta_i$; (ii) the persistence
of a low-energy suppression of the DOS (pseudogap) above $T_c$. As far as item (i) is
concerned, we show that inhomogeneity not only smears out the
coherence peak at $E_{peak}$, but it also to induce low-energy tails down to a 
quasiparticle gap $E_{gap}$ smaller than $E_{peak}$. On the other hand, the BCS optical response still displays a {\em sharp} optical absorption clearly located at $2E_{gap}$, and not at $2E_{peak}$. 
For what concerns item (ii), i.e. 
the presence of a suppression in the normal-state DOS, we show that it can lead to observable
deviations from the usual MB theory, which assumes a constant DOS. To account for this
latter effect we propose a modification of the MB formula, that compares very well
with the explicit theoretical calculations of the BCS response and with the experimental data
of Ref.\ \onlinecite{armitage15} at intermediate disorder. Finally, when we add the contribution of
collective modes we observe that the modifications to the BCS response manifest mainly
as an extra spectral weight below $2E_{gap}$. While at intermediate disorder this extra
absorption appears as a smearing of the BCS threshold at $2E_{gap}$, at
strong disorder the response of the collective modes resembles a peak at low frequency, which removes a considerable amount
of spectral weight from the superfluid response\cite{seibold_prl12}.  

The plan of the paper is the following.  In sec.\ \ref{sec:model} we introduce the model
and the basic definitions. In Sec.
\ref{sec:dos} we analyze the DOS of the system in the presence of disorder. With respect to previous work\cite{trivedi_prb01,nandini_natphys11} we clarify how disorder induces the formation of low-energy tails below the peak energy scale 
$E_{peak}$, leading to a lower value for the quasiparticle excitation threshold  
 $E_{gap}$. 
In Sec. \ref{sec:optcond} we report the results for the optical conductivity, and we
demonstrate that while the BCS response still displays a hard threshold at $2E_{gap}$, the collective modes induce additional  
low energy spectral weight. In Sec. \ref{sec:optcond} we analyze quantitatively the BCS response by providing a generalization of the  Mattis-Bardeen theory in order to reproduce the shape
of the normal-state DOS, which is strongly frequency-dependent
on the scale of the gap energy. We first demonstrate the validity of this 
generalization in the framework of the Hubbard model and then we show in Sec.\ \ref{sec:exp} that the theory gives also an excellent description of experimental data at intermediate disorder
as exemplified by the optical measurements from
Ref. \onlinecite{armitage15}. Finally, we compare our GI calculations with the experimental results on the most disordered films, where a substantial sub-gap absorption has been reported. Sec. \ref{sec:conc} contains our summary and the concluding remarks.

\section{Model and Formalism}\label{sec:model}
Our considerations are based on the attractive Hubbard model ($U<0$) with
local disorder                                                 
\begin{equation}\label{eq:ham}                                         
H=\sum_{ij\sigma}t_{ij}c^\dagger_{i\sigma}c_{j\sigma} + U\sum_{i}n_{i\uparrow}n_
{i\downarrow} +\sum_{i\sigma}V_i n_{i\sigma}                                      
\end{equation}                                                                    
which we solve in mean-field using the BdG transformation
\begin{equation}
\label{eq:bdg}
c_{i\sigma}=\sum_k\left[u_i(k)\gamma_{k,\sigma}-\sigma v_i^*(k)\gamma_{k,-\sigma
}^\dagger\right]\,.
\end{equation}
where the $u_i(k)$ and $v_i(k)$ variables are solution of the equations:
\begin{align}
\omega_k u_i(k)&=\sum_{j}t_{ij} u_j(k) + [V_i-\frac{|U|}{2}\langle n_i\rangle -
\mu] u_i(k) \nonumber \\ &+\Delta_i v_i(k)                              
 \label{eq1}\\                                                                  
\omega_k v_i(k)&=-\sum_{j}t^*_{ij} v_j(k) -
[V_i-\frac{|U|}{2}\langle n_i\rangle -\mu] u_i(k)\nonumber \\
&+\Delta^*_i u_i(k)\label{eq2}\,.
\end{align}
Here $\Delta_i$ and $\langle n_i\rangle$ represent the local SC order parameter and the local
density, respectively, defined by the self-consistent equations
\begin{eqnarray}
\label{op}
 \Delta_i&=&|U|\sum_k u_i(k)v^*_i(k) \\
 \label{nloc}
\langle n_i\rangle &=& 2\sum_k|v_i(k)|^2
\end{eqnarray}
The full sets of equations (\ref{eq1}-\ref{nloc}) are solved self-consistently giving
the configuration of the mean-field ground state in the presence of disorder. 
For simplicity only nearest-neighbor hopping $t_{ij}=-t$  is considered
in this work. The disorder variables $V_i$ are taken from a flat, normalized
distribution ranging from $-V_0$ to $+V_0$. Moreover, we will take units $e=\hbar=c=1$ in all the paper, unless explicitly stated. 

As it has been shown in several works before
\cite{trivedi_prb01,dubi_nat07,nandini_natphys11,seibold_prl12,lemarie_prb13,mandal_prl13,seibold_prb15}, 
the model (\ref{eq:ham}) describes already at mean-field level many features observed
experimentally in thin films of conventional SC near the SIT. In particular, the local 
SC order parameter $\Delta_i$ is strongly suppressed with increasing $V_0/t$, while the 
quasiparticle gap $E_{gap}$ (i.e. the minimun value of the quasiparticle energies $\omega_k$) saturates, indicating the formation of local, incoherent pairs. 
In addition, $\Delta_i$ segregates spontaneously by forming good SC regions embedded
in a poorly SC background. Here we investigate in detail how the modifications
introduced by this anomalous SC landscape in the DOS affect the optical conductivity, 
with the aim to understand how the emergence of several energy scales reflects in the
quasiparticle dynamics probed by optical spectroscopy.

In order to compute fluctuations on top of the (inhomogeneous) BdG ground state 
we evaluate dynamical  correlation functions
\begin{equation}
\chi_{ij}(\hat{A},\hat{B})=-i\int\!dt e^{i\omega t}\langle {\cal T} \hat{A}_i(t) \hat{B}_j(0)\rangle
\end{equation}
where in the following $\hat{A}$,$\hat{B}$ correspond
to either pair fluctuations, charge fluctuations or current operators,
i.e.
\begin{eqnarray*}
\delta\Delta_i &\equiv& c_{i\downarrow}c_{i\uparrow} - \langle c_{i\downarrow}c_{i\uparrow} \rangle \\
\delta\Delta_i^\dagger &\equiv& c^\dagger_{i\uparrow}c^\dagger_{i\downarrow} - \langle c^\dagger_{i\uparrow}c^\dagger_{i\downarrow} \rangle \\
\delta n_i^\dagger &\equiv& \sum_{\sigma}\left(c^\dagger_{i\sigma}c_{i\sigma} - \langle c^\dagger_{i\sigma}c_{i\sigma} \rangle\right) \\
j_i^\alpha &=& -i t \sum_\sigma\left\lbrack c_{i\sigma}^\dagger c_{i+\alpha,\sigma}
-c^\dagger_{i+\alpha,\sigma}c_{i\sigma}\right\rbrack 
\end{eqnarray*} 
and expectation values are evaluated with the BdG ground state. To compute the optical conductivity one needs in particular the current-current correlation function. For a given disorder configuration
the optical conductivity is given by
\begin{equation}\label{eq:oc}
  \sigma_{xx}(\omega))=-e^2\frac{1}{N}\sum_{ij}\frac{-t_i^x \delta_{ij}
    -\chi_{ij}(\hat{j_x},\hat{j_x})}{i(\omega+i\delta)}
\end{equation}
where $t_i^x$ is the kinetic energy between sites $R_i$ and
$R_{i+x}$. For definiteness we evaluate the response along
the x-direction. The results shown in the following are then
obtained by averaging over different disorder realizations.

As we discussed in the introduction, the optical conductivity depends crucially on the
level of approximation used to compute $\chi_{ij}(\hat{j_x},\hat{j_x})$. 
When the current response is evaluated at the level of the bare bubble we obtain the
optical conductivity in the BCS approximation, with the inclusion of the effect
of disorder on the quasiparticle excitations, described via the BdG
transformation (\ref{eq:bdg}).  The average over disorder restores the translational invariance of the system, but adds finite lifetimes to the energy levels in momentum space, in agreement with the general reasoning of the approach followed by 
Mattis and Bardeen\cite{mb},  with the difference that we do not make any
further assumption on the DOS in the normal or SC state. Once the vertex corrections are included all the contributions of the collective modes
to the optical response are taken explicitly into account at RPA level, and the
gauge-invariance of the theory is restored, as  required for the f-sum rule  to be
fulfilled\cite{schrieffer}.  Further technical details can be found
in Refs.\ \onlinecite{cea_prb14,seibold_prb15}.

\section{Density of states}\label{sec:dos}
\begin{figure}[htb]
\includegraphics[width=8.5cm,clip=true]{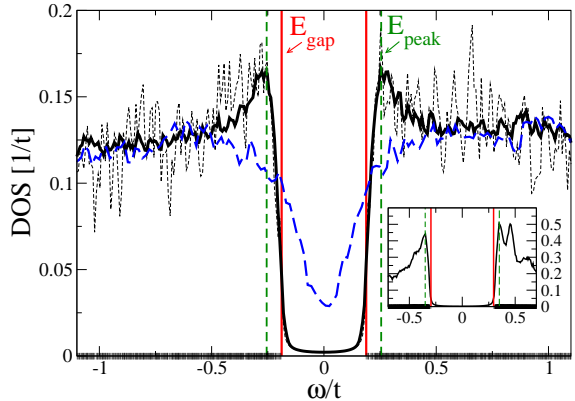}
\caption{Main panel: Density of states in the normal (blue) and
  superconducting state (black). The normal DOS is obtained 
  from an average over $150$ disorder configurations ($U/t=2$, $V_0/t=2$)
  of $24\times 24$
  systems. The SC DOS (solid, black) is the average over $30$ disorder configurations
  from $52\times 52$ systems whereas the black dashed curve is the DOS for a
  particular disorder realization. The eigenvalues of the latter system
  are indicated by bars on the frequency axis. Vertical lines
  indicate the position of the lowest eigenvalue (solid, red) and the
'coherence' peak feature (dashed, green).
  Inset: same for a system with disorder in the local interaction $U_i=U_0
  \pm \delta u_i$ with $U_0/t=2$ and $u_i$ is taken from a flat distribution
  between $-0.5 \le u_i \le +0.5$.
  In all cases the charge density is $n=0.875$.}
\label{fig1}
\end{figure}

Even though the overall behavior of the SC DOS for the model (\ref{eq:ham}) has been
already discussed previously\cite{trivedi_prb01,nandini_natphys11}, 
here we want to show more  specifically how disorder modifies its fundamental shape 
both in the normal and SC state with respect to the homogeneous case. 
As we mentioned in the Introduction, 
tunneling experiments on disordered superconductors deviate in several respects from the usual BCS paradigm. 
First, a Altshuler-Aronov (AA) type correction to the DOS \cite{altshuler} is found, in the form of an 
extended dip around the Fermi level already at high temperatures\cite{mondal10,chand12}. 
Second, a pseudogap occurring at the energy scale of the SC gap survives above $T_c$, in 
a range of temperatures that increases for increasing disorder\cite{sacepe10,mondal10,sacepe11,chand12,kam13}.
Fig. \ref{fig1} reveals a similar pseudogap feature in the normal state DOS, shown with a blue dashed line. 
In the present case this gap is mainly caused by the unconstrained variation
of the densities in the presence of an attractive onsite interaction.
In fact, the local chemical potential at site $R_i$ is given by
$E_i=V_i+U n_i/2$ so that sites with $E_i$ just below (above) the
Fermi energy tend to increase (lower) the local density in order
to lower the total energy. This redistribution of charge is limited
by the concomitant loss in kinetic energy and leads to the pseudogap
in the normal state.

The black thick  line in the main panel of Fig. \ref{fig1} shows the DOS in the superconducting state, 
obtained as an average over $10$ disorder configurations for large systems with $52 \times 52$ sites. 
The black thin dashed line is the DOS for a particular disorder realization.
As can be seen both curves agree in the evolution of spectral weight from inside the
gap towards the maximum DOS located at $E_{peak}$, which resemble coherence peak features.
The eigenvalues of the BdG matrix (corresponding to the black dashed DOS)
are shown as bars on the frequency axis. Clearly the DOS between the
lowest (absolute) eigenvalue (i.e. the hard gap $E_{gap}$)
and the 'coherence peak' is determined by a large set of eigenvalues
which induce a tail feature in the DOS. For this reason, the
tail is also not related to the small $\eta=0.005t$ which determines
the width of the lorentzians in our evaluation of the DOS.
Thus, in contrast to the textbook profile of the BCS DOS, where the sharp coherence peaks mark
also the onset of the hard gap, here two separate energy scales occur. 
In general, the hard gap $E_{gap}$ measuring the minimum energy required to create a
single-particle excitation is lower than the value $E_{peak}$, where the spectral weight
removed from low energy piles up giving rise to a smeared coherence peak. 
Notice that the separation between these two energy scales is similar to what has been discussed before 
in models with inhomogeneous distribution of the coupling constants\cite{armitage15,feigelman12}, based on the Larkin-Ovchinnikov original suggestion\cite{larkin71}. Indeed, the same result could be reproduced in our approach by
introducing disorder in the local interaction $U_i$ instead of the local chemical potential.  The result for   $U_i/t=2.0/t \pm \delta u_i$, with $u_i$ taken from a flat distribution 
  between $-0.5 \le u_i \le +0.5$a, is shown in the inset to Fig. \ref{fig1}. Clearly, the main features are
  similar to those of the main panel, namely a large set of eigenvalues which determines the smooth evolution
  of the DOS between the hard gap and the coherence peaks.

\section{Optical conductivity}\label{sec:optcond}
\begin{figure*}[htb]
\includegraphics[width=8.5cm,clip=true]{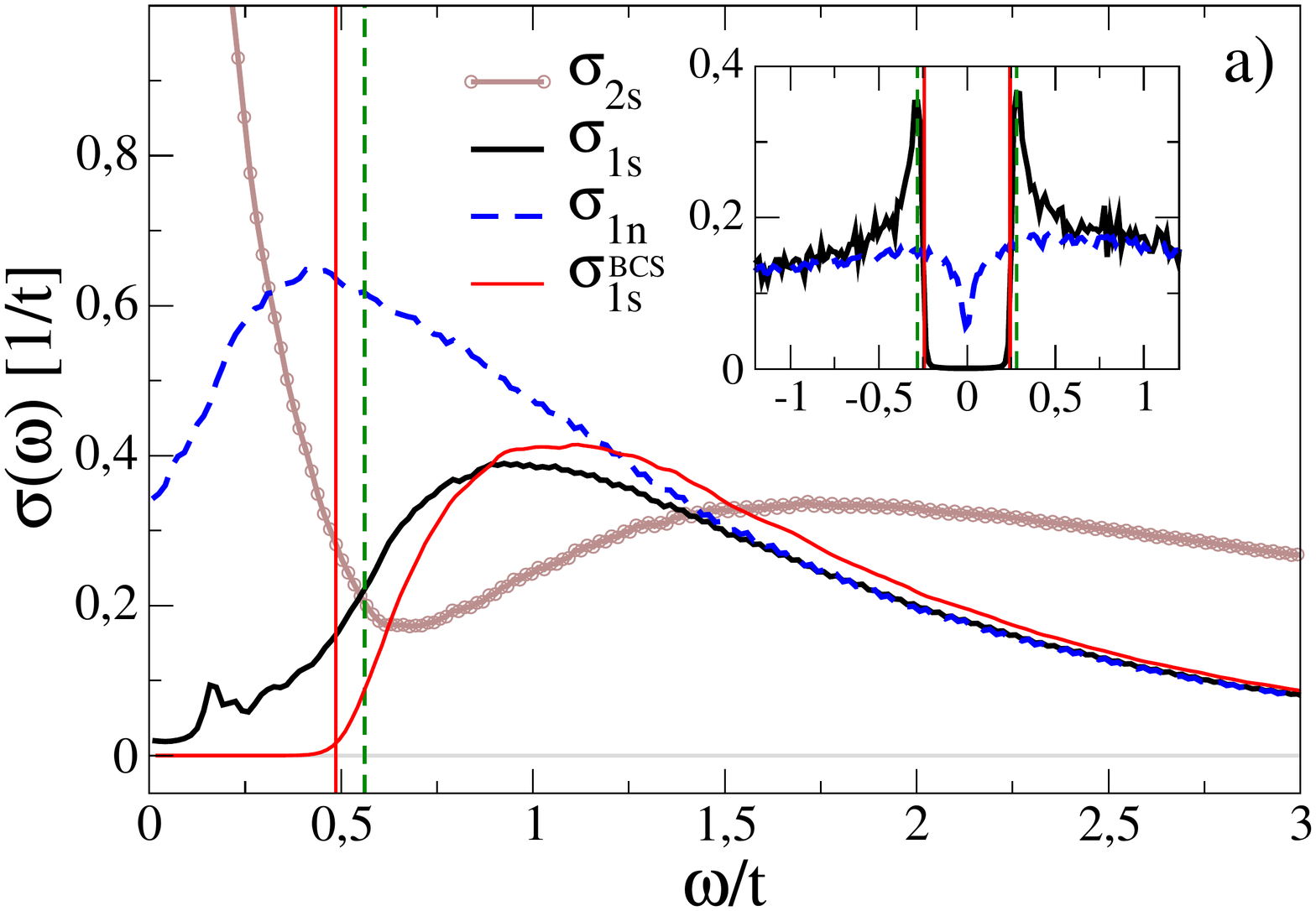} 
\includegraphics[width=8.5cm,clip=true]{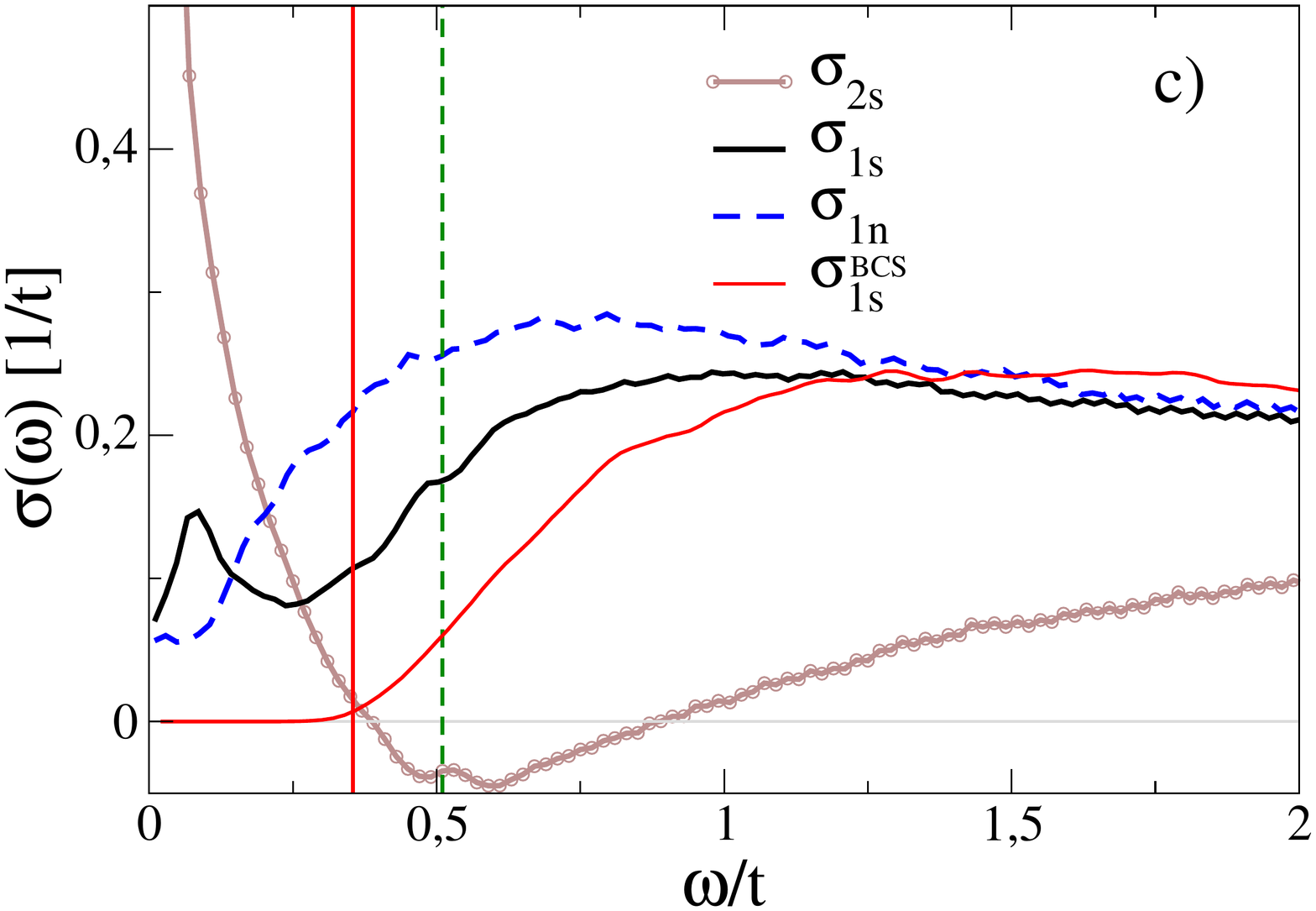}
\includegraphics[width=8.5cm,clip=true]{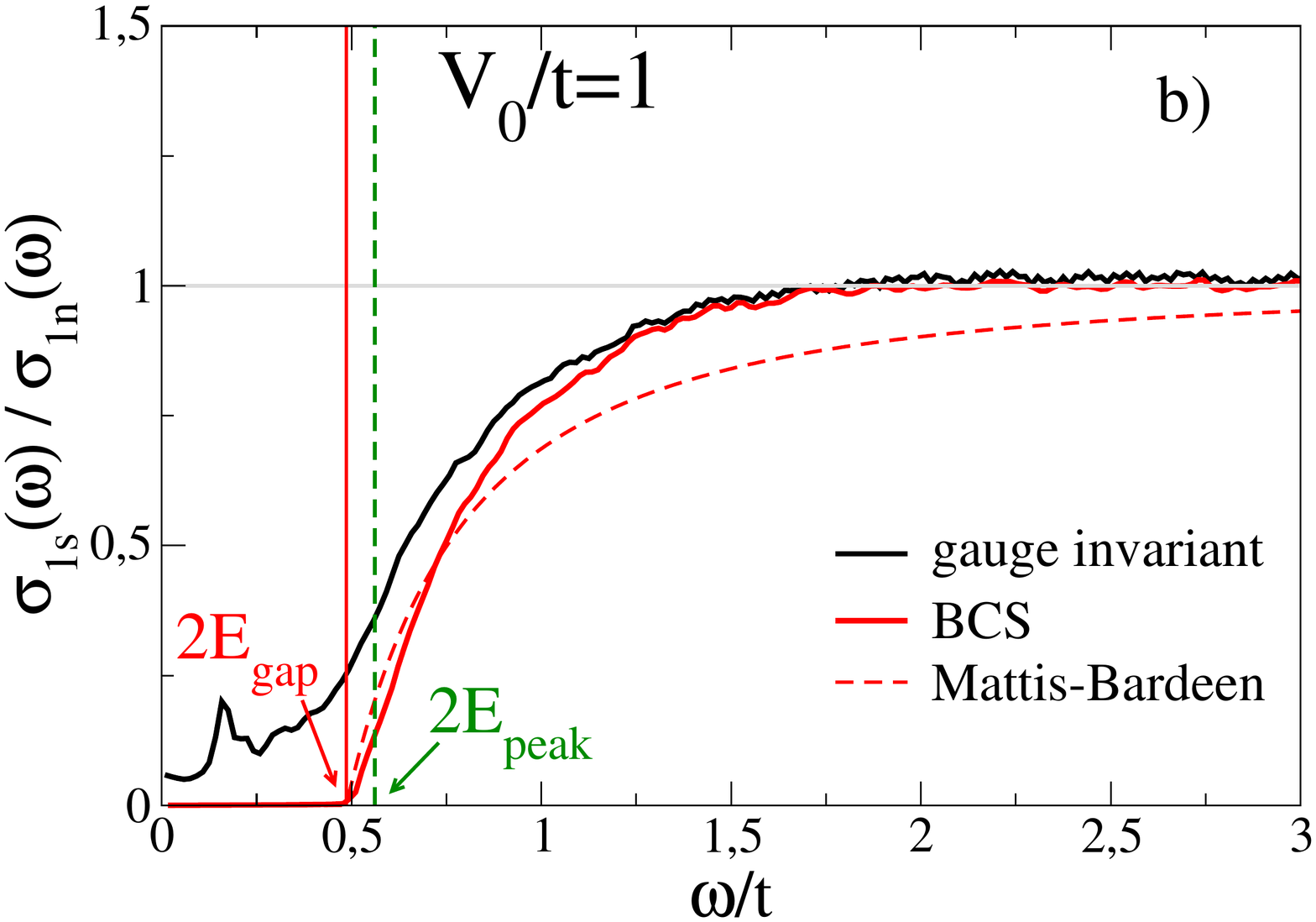}
\includegraphics[width=8.5cm,clip=true]{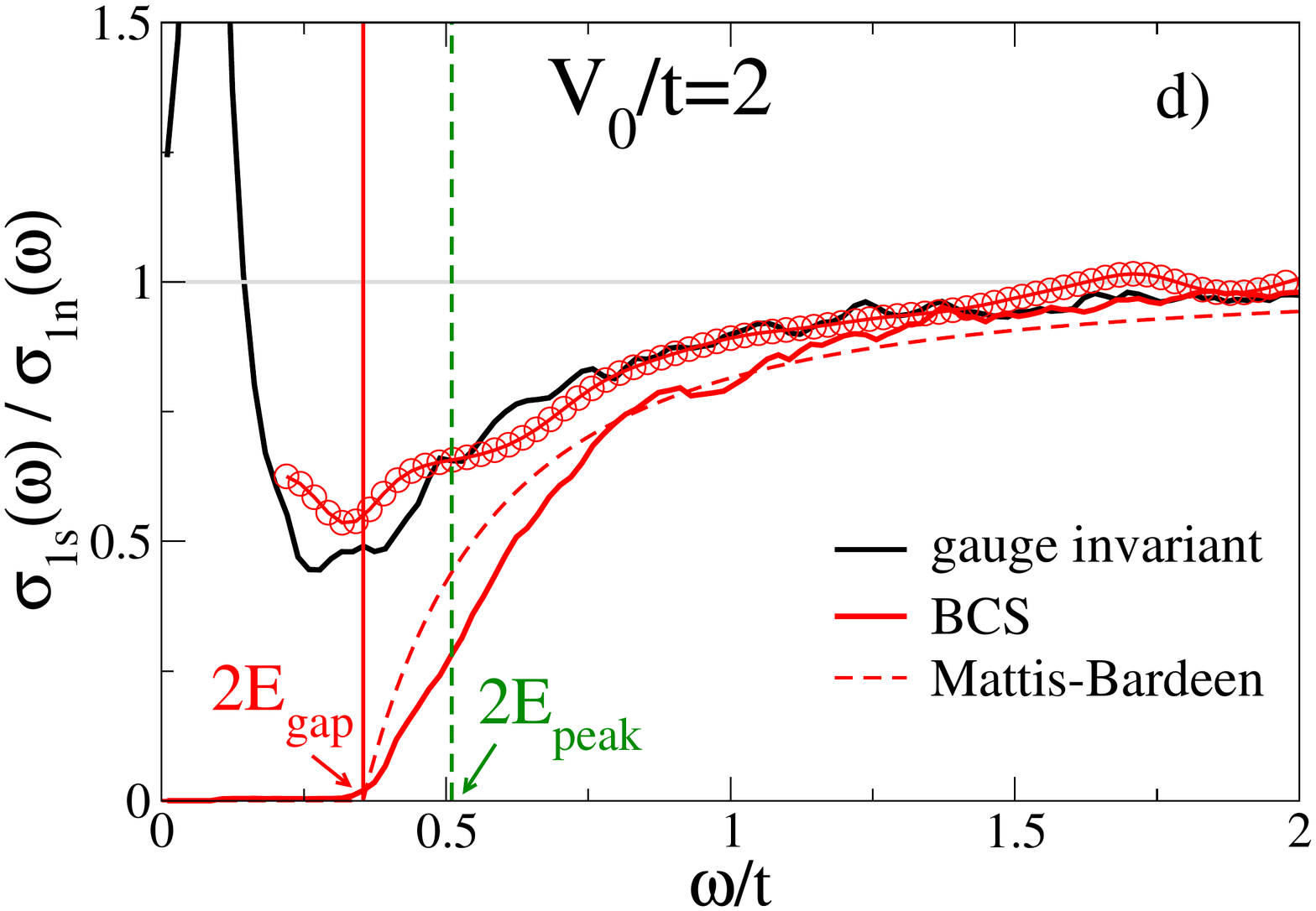}
\caption{Optical conductivity at $V_0/t=1$ (panels a,b) and at $V_0/t=2$ (right panels c,d). 
The remaining parameters are: $U/t=2$, $n=0.875$, $24\times 24$ lattice.
Panels (a) and (c) show the real (solid, black) and imaginary (dotted, maroon) part of 
the optical conductivity in the SC sate and $\sigma_1$ in the normal state (dashed, blue) 
evaluated as an  average over $50$ disorder configurations within the gauge-invariant RPA 
approach. For comparison, the real part of the conductivity in the SC state within
the BCS approximation is also shown by the red solid line.  The vertical green and red 
lines denote $2E_{peak}$ and $2E_{gap}$, respectively, as extracted from the DOS. The DOS for
the case $V_0/t=2$ is the one shown in Fig.\ (\ref{fig1}), while the one for $V_0/t=1$ is 
shown explicitly as an inset of panel (a). Panels (b) and (d) show the ratio between the 
real part of the optical conductivity computed in the SC and normal state. The red curve 
reports the corresponding result in the BCS limit and the dashed line is the Mattis-Bardeen
fit done using $2E_{gap}$ as the optical threshold. For comparison, panel (d) also 
shows experimental data from Ref. \onlinecite{armitage15} for the most disordered sample (T$_c=3.8 K$)
and we have scaled the energy by $t=0.8$ THz.
Notice that for the BCS case also 
$\sigma_{1,n}$ has been computed at the bare-bubble level, leading to a normal-state conductivity 
slightly different from the gauge-invariant result reported in the upper panels.}
\label{fig2}
\end{figure*}

Once the modifications of the DOS in the normal and SC state due to an inhomogenous SC ground state
have been established, let us now investigate how disorder affects  the optical conductivity. 
Denoting by $\chi_{1,2}(\omega)$ the real/imaginary part of the disorder average of 
$1/N\sum_{n,m}\chi_{nm}(\hat{j_x},\hat{j_x})$ (i.e. the $q=0$ component),
and by $\langle t_x\rangle$ the disorder average of the mean kinetic energy along
the x-direction, one finds for the real and imaginary part of $\sigma(\omega)$ 
\begin{eqnarray}
\label{sigma1}
\sigma_1(\omega)&=&\pi D_s\delta(\omega) - \frac{\chi_2(\omega)}{\omega}, \\
\sigma_2(\omega)&=& \frac{-\langle t_x\rangle + \chi_1(\omega)}{\omega}\label{eq:s2}\,.
\end{eqnarray}
where the superfluid stiffness $D_s$ is given by $D_s=\lbrack -\langle t_x\rangle + \chi_1(\omega=0) \rbrack$. 
Note that Fig. \ref{fig2} only shows the regular part of
$\sigma_1(\omega)$, i.e. without the superfluid delta-like contribution of Eq.\ (\ref{sigma1}). 
It can be seen from Eq. (\ref{eq:s2}) that the stiffness is also
obtained as the limit
$D_s =\lim_{\omega\to 0} \omega \sigma_2(\omega)$. 

In a perfectly clean system the RPA current-current correlation function at finite frequency
vanishes, so that $D_s$ equals the so-called diamagnetic term $D_d\equiv-\langle t_x\rangle$, 
which coincides with $n/m$ in a system with parabolic band dispersion. In this case
$\sigma_1(\omega)=\pi D_s\delta(\omega)$ and $\sigma_2(\omega)=D_s/\omega$.\cite{nota_sigma} Disorder changes
this picture in two ways. First, similarly to what happens in the normal state, 
single-particle excitations acquire a finite lifetime
broadening $\Gamma$ in momentum space, making finite-frequency absorption possible, 
as encoded in a finite value of $\chi_2(\omega)$ in Eq.\ (\ref{sigma1}). Since by
Kramers-Kronig $\chi_1(0)=\int d\omega \chi_2(\omega)/\omega$  the finite-frequency
absorption results in a reduction of the stiffness $D_s$ with respect to the diamagnetic
term $D_d$. When $\chi(\omega)$ is computed at BCS level one obtains the suppression
of $D_s$ due to MB theory of the order of $D_s\sim D_d (\Delta/\Gamma)$. Notice that the 
optical sum rule is always satisfied, since the total integrated spectral weight 
\begin{equation}
\int_0^\infty d\omega \sigma_1(\omega)=\pi D_d
\end{equation}
still scales with $D_d$. Second, once that collective modes are included via the vertex corrections one expects
a further reduction of $D_s$, as it has been pointed out already in 
Ref. \onlinecite{seibold_prl12} by direct computation of the stiffness within a 
gauge-invariant approach.  In particular, it turned out that $D_s$ can be reduced by orders
of magnitude with respect to its BCS counterpart in the presence of strong disorder. 
As a consequence the imaginary part of $\sigma(\omega)$ can even become {\em negative} at 
small energies, a feature which is also observed experimentally \cite{armitage15} for the 
strongly disordered samples. In addition, the suppression of the stiffness will be 
compensated, due to the optical sum rule,  by an extra absorption with respect to the BCS or MB bare-bubble estimate.

All these features are correctly captured by the BCS and the full gauge-invariant RPA calculations of 
the optical conductivity shown in Fig.\ \ref{fig2} for two values of disorder. In the 
upper panels we report the absolute values of $\sigma_1$ and $\sigma_2$ in the SC state, 
along with $\sigma_1$ in the normal state. As one can see, already at intermediate disorder 
$\sigma_1(\omega)$ shows a suppression at low-energy in the normal state, that is specular 
to the analogous suppression of the DOS discussed in Fig.\ \ref{fig1} (see also the 
DOS at $V_0/t=1$ shown in the inset of Fig.\ \ref{fig2}). When one enters the SC 
state $\sigma_1(\omega)$  is further suppressed due to the opening of a full SC gap. 
The optical spectral weight is then transferred to the delta peak at zero frequency (not shown), 
and manifests in a finite imaginary response $\sigma_2(\omega)$ that diverges as $D_s/\omega$ 
at low frequencies. 

To establish a closer connection with the usual MB results we report in the lower panels of Fig.\ \ref{fig2}
the ratio between $\sigma_{1,s/n}$ in the SC and normal state, in the case of a BCS (red line) 
or full gauge invariant (black line) calculation.  Notice that despite the strong 
inhomogeneity of the underlying SC state\cite{trivedi_prb01,nandini_natphys11,seibold_prb15} 
the BCS ratio $\sigma_{1,s}/\sigma_{1n}$ shows a hard threshold developing at twice the 
quasiparticle excitation gap $E_{gap}$, and not at twice the  coherence-peak scale 
$E_{peak}$, as it has been sometimes argued in the analysis 
of the experimental data\cite{frydman_natphys15}. This is particularly evident 
in Fig.\ \ref{fig2}d where the separation between the two scales is larger. The 
persistence of a hard gap in the BCS response suggests to compare the BCS calculations 
with the outcomes of the standard MB formula, with $\Delta$ replaced by $E_{gap}$. As can 
be seen  in Fig.\ \ref{fig2}b,d the overall trend of the BCS result is captured by the 
standard MB formula, even though the MB fit overestimates the absorption close to the 
onset, whereas for large $\omega$ it stays slightly below the BCS curve.  As we will 
discuss in more details in the next section, these discrepancies can be attributed to the 
assumption of a constant normal-state DOS, implicit in the MB approach. 

When the contribution of the collective modes is considered the extra optical absorption 
becomes finite even below $2E_{gap}$, with an intensity and shape which depend on the 
disorder level. At intermediate disorder, Fig.\ \ref{fig2}b, the collective modes are 
responsible for the emergence of a tail below the quasiparticle optical gap, that 
strongly resembles the recent measurements of Ref.\ \onlinecite{armitage15}. In this respect, 
our results share analogies and differences with respect to the interpretation proposed recently in 
Ref. \onlinecite{armitage15}. Here the authors 
argue that the existence of quasiparticle states between the true optical gap, i.e. $2E_{gap}$, and the scale $2E_{peak}$
identified by the maximum in the DOS, can be captured by means of an inhomogeneity-induced smearing of the conventional
MB response starting at $2E_{peak}$. The inhomogeneous MB fit in Ref. \onlinecite{armitage15} is derived from a
distribution of the coupling constant \cite{larkin71}
so that $2E_{peak}$ is still a meaningful scale related to the average SC gap.
Our results, which instead are derived from local onsite disorder, confirm the identification of the optical gap
with $2E_{gap}$, but they also show that the scale $2E_{peak}$ does not play any role in the BCS response, which displays a
MB-like sharp optical onset at $2E_{gap}$.
The extra absorption below the $2E_{gap}$ threshold (which from an experimental perspective
hampers the identification of this energy scale \cite{armitage15})
is due instead to collective modes, whose 
contribution can even become visible well above it at stronger disorder, 
see Fig.\ \ref{fig2}d. As we shall see below, our view is further supported by 
the direct comparison of the BCS calculations with experimental data in samples at 
intermediate disorder,  that can be very well reproduced within a modified MB scheme with 
the only exception of a small sub-gap tail.

\section{Effect of a pseudogap in the Mattis-Bardeen formula}\label{sec:mb}
The MB theory\cite{mb} starts from the Pippard's expression describing the non-local relation between the 
current density  ${\bf j}({\bf r},t)$ found in response to an applied frequency dependent vector potential 
${\bf A}({\bf r}',\omega)$. In order to account for disorder, the idea is
to multiply the BCS kernel 
$K({\bf r}-{\bf r}',\omega)$, relating these two quantities, with a factor
$\exp[-|{\bf r}-{\bf r}'|/(2l)]$ where $l$ is the mean-free path. 
 For strong disorder and in the Pippard limit $\lambda\ll \xi_0$ this simplifies the integration over  ${\bf r}'$, leading to a rather simple expression for the complex optical conductivity. Later on it has been shown by Nam\cite{nam67} that the same result holds also in the London limit ($\lambda\gg\xi_0$) as long as the impurity scattering is weakly frequency dependent. Nonetheless, in both cases one assumes that the DOS in the normal state is constant over the energy scale of the SC gap, giving a constant $\sigma_1$ above $T_c$.  
However, as discussed in the previous Sections, the DOS of a strongly-disordered superconductor is 
expected to develop a  pseudogap already in the normal state. 
This feature, that is reproduced by our results already at the level of the BCS approximation, has been also observed in 
the tunneling spectra of several systems\cite{mondal10,kam13,roditchev13}. 

In order to reproduce the correct BCS result we rederived the MB formula  by explicitly keeping any energy dependence of the normal-state DOS $N(\omega)$ on the scale of the SC gap. In the following we will also assume a DOS 
 symmetric around the chemical potential $N(\omega)=N(-\omega)$, while the general result is given by Eq. \ref{eq:amb} in appendix \ref{sec:a1}.
For the symmetric case one finds
\begin{eqnarray}
  \frac{\sigma_{1s}}{\sigma_{1n}}&=&\frac{I(\omega)}{\Gamma(\omega)} \label{eq:mbf} \\
\Gamma(\omega)&=& \frac{1}{\omega}\int_{-\infty}^\infty \! d\epsilon N(\epsilon) 
N(\epsilon+\omega)\left\lbrack f(\epsilon)-f(\epsilon+\omega)\right\rbrack\label{eq:gw} \\
   I(\omega)&=& \frac{2}{\omega} \int_\Delta^\infty\!dE\,\frac{f(E)-f(E+\omega)}
   {\epsilon_1\epsilon_2} \label{eq:iw}\\
   &\times&\left\lbrack E(E+\omega) +\Delta^2\right\rbrack 
   N(\epsilon_1)N(\epsilon_2) \nonumber\\
  &-& \frac{\Theta(\omega-2\Delta)}{\omega}\int_{\Delta-\omega}^{-\Delta}\!dE\,\frac{1-2f(E+\omega)}
  {\varepsilon_1\varepsilon_2}\nonumber\\ &\times&\left\lbrack E(E+\omega) +\Delta^2\right\rbrack 
  N(\epsilon_1)N(\epsilon_2)\nonumber\\
  \label{eps1}
   \epsilon_1 &=& \sqrt{E^2-\Delta^2} \\
   \label{eps2}
    \epsilon_2 &=& \sqrt{(E+\omega)^2-\Delta^2}
\end{eqnarray}
where  the functions $\Gamma(\omega)$ and $I(\omega)$ correspond, up to a constant, to the conductivities in the normal and SC state, respectively. 
In fact, $\Gamma(\omega)$ approaches
a constant for a constant DOS
$N(\omega)=const.$, as it is the case for the normal-state conductivity in the standard MB theory, where the broadening of the Drude peak occurs on the scale of a scattering rate much larger than  $\Delta$. 

Concerning the imaginary part of $\sigma(\omega)$ the 'standard' way in deriving this
quantity is via a continuation of the square-root functions in the complex plane, i.e. 
in the quantity $\epsilon_{2}$ in Eq.\ (\ref{eps2}). 
However, this procedure in the present case is quite cumbersome since $\epsilon_{2}$ 
also appears in the frequency-dependent DOS. 
Thus, we find it more convenient to calculate $\sigma_2(\omega)$ via a Kramers-Kronig
transformation of $\sigma_1(\omega)$ as given by Eq.\ (\ref{eq:mbf}). To account also for 
the delta peak of $\sigma_1(\omega)$ at zero frequency, see Eq.\ (\ref{sigma1}),  
we add to $\sigma_2$ a $1/\omega$ term whose strength can be obtained from the standard
MB expression for $\sigma_2(\omega)$ in the
limit $\omega\to 0$ (see Eq.\ (\ref{sigma2mb})) where the DOS at the chemical potential $N(0)$ can be taken
in front of the integral,
\begin{eqnarray}
\lim_{\omega\to 0}\frac{\omega\sigma_2(\omega)}{\sigma_{1n}}&=&\frac{n_s e^2}{m\sigma_{1n}}\label{eq:ns} \\
&=&\frac{N^2(0)}{\Gamma(0)}\lim_{\omega\to 0}
\int_{\Delta-\omega}^{\Delta}\!dE\,[1-2f(E+\omega)] \nonumber\\
&\times& \frac{E(E+\omega) +\Delta^2}{\sqrt{\Delta^2-E^2}\sqrt{(E+\omega)^2-\Delta^2}}\\
 &=& \pi \Delta \label{eq:ds}
\end{eqnarray}
and the last equality holds in the limit of zero temperature.

\begin{figure}[htb]
\includegraphics[width=8.5cm,clip=true]{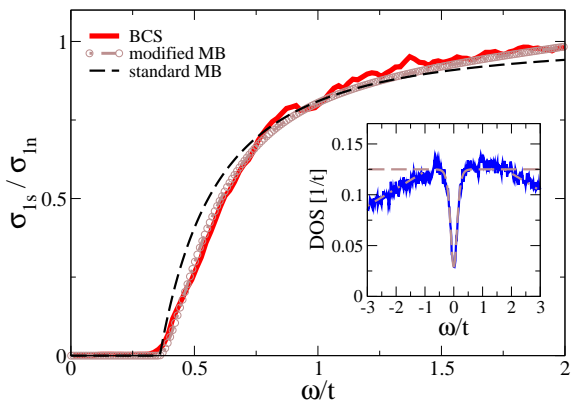}
\caption{Ratio between optical conductivities in the
  superconducting and normal state for the Hubbard model at $V_0/t=2$ (as in 
  Figs. \ref{fig1} and \ref{fig2}d) in the BCS approximation (red line), compared to 
  the prediction of the standard MB theory (dashed black line) and of the modified MB
   formula (\ref{eq:mbf}) (brown symbols) with $\Delta/t=0.36$. Inset: fit of the numerical results for the 
   normal-state DOS  of the Hubbard model. Near the Fermi level we used  the approximated
formula (\ref{eq:dos}) with $\alpha=0.0156/t,\beta=0.1095/t, \Omega=0.1968t$, 
while away from it we used a linear extrapolation at large energies. }
  \label{fig3}
\end{figure}

To test the validity of  Eq. \ref{eq:mbf} we first compare it with the results obtained for the 
optical conductivity at the BCS level, shown by the red solid line in Fig. \ref{fig2}b,d.
To model the energy-dependent DOS we use the following approximation for the normal-state DOS
\begin{equation}\label{eq:dos}
  N(\omega)/N(\infty)=\alpha+\beta\tanh^2(\frac{\omega}{\Omega})\,.
\end{equation}
As one can see in the inset to Fig. \ref{fig3}, Eq. (\ref{eq:dos}) yields a very good approximation
to the disorder averaged BdG DOS around the chemical potential. However, since the
latter deviates from a bare constant at larger energies, we have in addition
also implemented a DOS linear in frequency beyond the pseudogap, 
as shown by the dashed-dotted lines in the inset to Fig. \ref{fig3}.

The modified absorption due to the presence of an energy-dependent DOS 
is compared with standard MB theory in the main panel of Fig. \ref{fig3}. 
As one can see, accounting properly for the energy dependence of the normal-state excitations 
yields an excellent description of the BCS results within the modified MB scheme (\ref{eq:mbf}). 
First, the DOS suppression near the Fermi level allows us to correctly capture the onset of 
absorption close to the hard gap of the DOS, that is overestimated within the
standard MB formula. At larger frequency the linear trend observed in the DOS reflects in 
a saturation of the BCS result faster than the standard MB prediction at
$\omega/t\approx 1$, that is again well captured by our modified MB scheme. 

\begin{figure}[htb]
\includegraphics[width=8.5cm,clip=true]{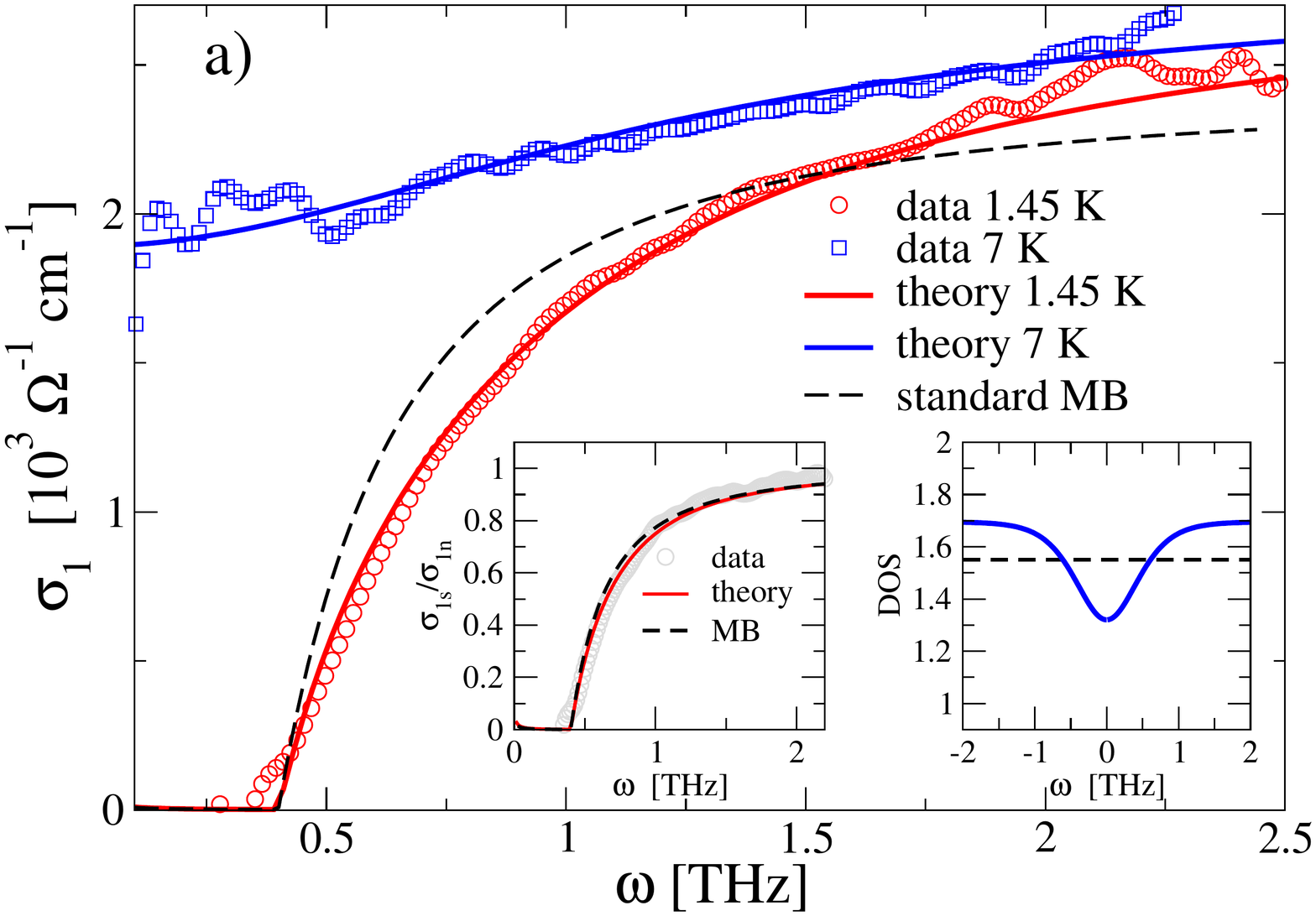}
\includegraphics[width=8.5cm,clip=true]{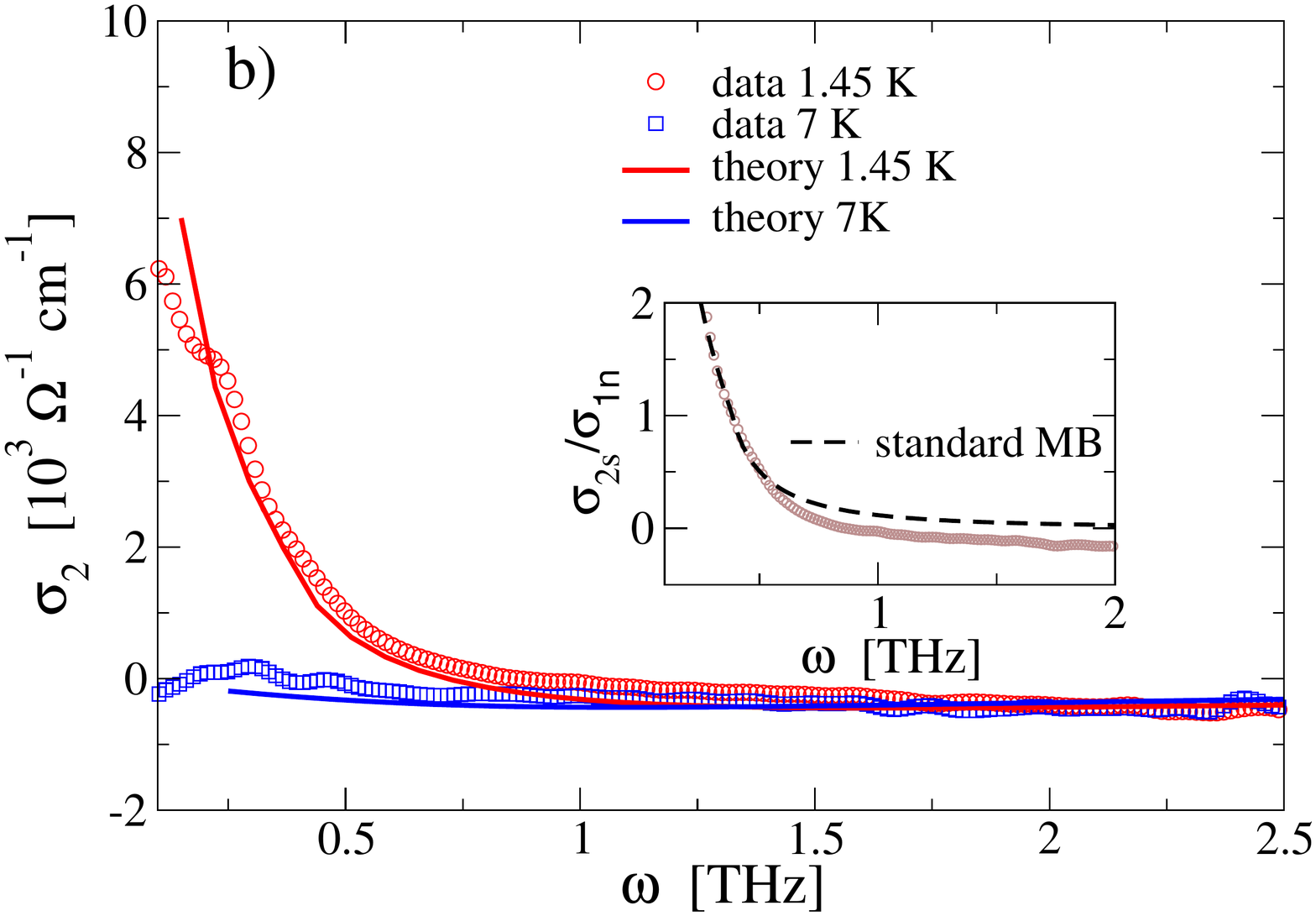}
\caption{Real (a) and imaginary (b) part of the optical conductivity for 
a NbN sample with $T_c=6.2 K$. Data at  $T=7 K$ and $T=1.45 K$  are shown
by squares (blue) and circles (red); the corresponding fits from the modified MB 
theory Eq.\ (\ref{eq:mbf}) are the solid lines, while the dashed black line is the 
standard MB formula with a constant DOS. In both cases we used $\Delta=0.2 THz$.The left 
inset of panel (a) and the inset (b) shows the ratio between the $T=1.45$ and $T=7 K$ data 
(symbols) compared with conventional and modified Mattis-Bardeen theory. Notice that 
while in the ratio $\sigma_{1s}/\sigma_{1n}$ the effect of the DOS disappear, making 
the standard MB formula apparently good, the ratio $\sigma{2s}/\sigma{1n}$ cannot be 
well reproduced by the standard MB approach. The right inset of panel shows the DOS 
used to reproduce the normal-state conductivity, with parameters $\alpha=1.32$, 
$\beta=0.37$, $\Omega=0.58 THz$ in Eq. \ref{eq:dos}. 
 Experimental data by courtesy of the authors of Ref.\ [\onlinecite{armitage15}]}
\label{fig4}
\end{figure}

\section{Comparison with the experiments}\label{sec:exp}
Once clarified the role of the inhomogeneous SC background and of the pseudgap on the BCS 
and full GI optical conductivity let us now compare our results with the experimental data 
from Ref.\ \onlinecite{armitage15}, where a systematic investigation of the role of 
disorder in several NbN samples has been provided. We first focus on a sample in the 
intermediate disorder regime ($T_c=6.2 K$), where one finds a  pseudogap behavior in the 
normal state, evidenced by a depression of $\sigma_{1n}$ at low 
frequencies (see Fig.\ \ref{fig4}a), but without the emergence of significant subgap features. 
According to the previous discussion, we expect that in this regime the optical conductivity 
can be well reproduced by the BCS calculation. To account also for the effects of an 
energy-dependent DOS we then apply the modified MB formula (\ref{eq:mbf}). 
We determine the parameters entering the DOS Eq. (\ref{eq:dos})
by minimizing the differences between experimental and theoretical $\sigma_{1}(\omega)$
[from Eqs. (\ref{eq:gw}, \ref{eq:iw})] 
in the normal and superconducting state, respectively.
The resulting DOS is reported in the lower right inset to Fig. \ref{fig4}a
with a depletion at the chemical potential of $\beta/(\alpha+\beta)=22\%$  
and width $\Omega=0.58 THz$.
Fig. \ref{fig4} demonstrates that for the real part of for $\sigma(\omega)$ this approach
leads to an excellent agreement with experimental data both in the
superconducting ($T=1.45 K$) and  normal ($T=7 K$) state, respectively.
On the other hand, standard Mattis-Bardeen theory with a constant DOS (dashed line
in Fig. \ref{fig4}) overestimates the recovery of spectral weight in the superconducting
state and would of course predict a constant $\sigma_{1n}$ in the normal state. 

The fit for the imaginary part $\sigma_2(\omega)$ is shown in panel (b) of Fig. \ref{fig4}.
Below $T_c$ the contribution arising from the $1/\omega$ part, calculated from Eq. (\ref{eq:ds}), 
depends sensitively on the value of the DOS at the chemical potential $N(0)$. Since the  experiment strongly looses signal below $\sim 0.3 THz$ it is difficult
to estimate this value from the optical data. Despite this uncertainty the fit
from our extended Mattis-Bardeen theory gives a good account of the data, both below
$T_c$ and in the normal state. 
In Fig. \ref{fig4}a,b we also show the effect of taking the ratio 
$\sigma_{1s}/\sigma_{1n}$ in analyzing the data. As one can see, while for the
 real part the errors in the standard MB formula cancel out each other in the
fit of $\sigma_{1s}/\sigma_{1n}$, for the imaginary part the usual MB approach fails
 completely to recover the frequency dependence of $\sigma_{2s}/\sigma_{1n}$ above the optical gap. 
The analysis for a second, less
disordered, sample is reported in appendix \ref{sec:apb}.

At stronger disorder the microwave absorption deviates substantially from the MB, and even
 our modified MB scheme cannot reproduce the experiments, since the effect of collective 
 modes should be included. In this regime our model system Eq.\ (\ref{eq:ham}) does not 
 have enough flexibility to reproduce separately both the normal-state pseudogap of the 
 DOS and the anomalous sub-gap absorption below $T_c$. Nonetheless, by tuning only the 
 value of the hopping, which fixes the energy units in the Hubbard model, we can give a 
 good account for the behavior of $\sigma_{1s}/\sigma_{1n}$, as shown in panel (d)
 of Fig.\ \ref{fig2}. 
 This result is particularly striking, considering the lack of Coulomb interactions in 
 our model. Indeed, as it has been discussed in Ref. \onlinecite{cea_prb14}, the extra spectral 
 weight at strong disorder is due mainly to the contribution of phase fluctuations, which 
 become optically active due to the
presence of disorder\cite{cea_prb14,stroud_prb00,trivedi_prx14}. Since in our model the phase mode is a 
sound, the extra absorption manifests mainly at low frequencies. In real systems one could 
expect that long-range Coulomb interactions pushed at least in part this extra absorption
 at higher energy. However, the surprisingly good qualitative and quantitative agreement 
 between our full GI calculations and the experiments shown in Fig.\ \ref{fig2} suggests 
 that this effect is not so severe. We expect that this is due both to the small value of the stiffness and to the strong inhomogeneity induced by disorder, that makes the optically-active phase modes rather local and therefore less sensitive to the Coulomb interaction.


\section{Conclusions}\label{sec:conc}
In this work we analyzed several key aspects which must be taken into account to correctly interpret the optical measurements in disordered
SC films from weak to strong disorder, in order to account for significant deviations from the conventional MB theory. As we discussed in detail, in general the optical response can be computed in the bare-bubble or BCS approximation, that corresponds to include only the effects of quasiparticle excitations,  or beyond that by adding the effect of collective modes via vertex corrections. The MB theory corresponds to the BCS approximation in the presence of disorder, included via the introduction of a finite length scale setting the correlation between two plane-wave states. As a 
consequence, within the MB approach disorder does not influence the DOS, which is a constant in the normal state and reduces to the usual clean BCS form in the SC state, with two sharp coherence peaks at the gap value and zero available state below it.  In our paper we have critically analyzed these assumptions by evaluating
the optical response for systems where disorder is implemented by a random on-site potential, and the optical response is computed both in the BCS approximation and by including vertex corrections at RPA level. 

We first analyzed the BCS optical response, by making a direct connection with the average DOS of our disordered system, which shows itself significant deviations with respect to the clean case. Indeed, the self-organized formation of SC islands induced by disorder suppresses considerably the sharp coherence peaks at $E_{peak}$ and generates tails extending down to a lower hard gap $E_{gap}$. This effect was already known to occur for inhomogeneities in the SC coupling parameter\cite{larkin71,feigelman12}. However, our model of disorder seems more appropriate to describe the experimental observation\cite{kam13} that the formation of nanoscale SC islands in the
disordered films is {\it not} related to the atomic scale inhomogeneity, which would also induce a related 'atomic scale' modulation of
the coupling constant.  
Despite the existence of two separate energy scales $E_{gap}<E_{peak}$ in the DOS, the BCS response still displays a hard optical absorption at $2E_{gap}$. This effect can be well captured by the MB formula provided that this energy scale, and not $2E_{peak}$ as usually assumed\cite{frydman_natphys15,armitage15}, is identified with the parameter $2\Delta$ in the MB formula.  We have shown (cf. Figs. \ref{fig4},\ref{fig5}) that this identification strongly improves the analysis of the optical data for less disordered samples within MB theory. Further improvement can be achieved by implementing the presence of a pseudogap feature in the normal-state DOS, i.e. a  DOS frequency-dependent on the scale of  $E_{peak}$ (cf. Figs. \ref{fig4},\ref{fig5}). We have shown that this effect can be included within the MB scheme via the formulas Eqs. (\ref{eq:mbf}-\ref{eps2}), that could be
helpful for experimental data analysis, allowing for a consistent fit of both $\sigma_{1n}$ and $\sigma_{1s}$ separately. In fact,
as shown in Fig. \ref{fig5}, a MB fit of $\sigma_{1s}/\sigma_{1n}$ alone can
artificially yield a reasonable fit although $\sigma_{1n}$ displays strong
variations at low energy. The consistent procedure suggested by
our analysis would be to first extract the frequency-dependent DOS via a fit of the normal-state optical conductivity and then to implement it in the fit of optical data in the SC state, which then provides the value of $2E_{gap}$. 

At stronger disorder the BCS approximation fails, since collective modes become optically active. We have shown that their inclusion within a gauge-invariant RPA scheme removes the sharp onset of absorption at $2E_{gap}$, building up additional spectral weight below the quasiparticle contribution.  From an experimental point of view this hampers 
the determination of $2E_{gap}$ which in some cases may be still extracted from
an extrapolation of an extended MB fit [Eqs. (\ref{eq:mbf}-\ref{eps2})] and
a concomitant analysis of the tunneling data. As discussed in Ref.\ [\onlinecite{cea_prb14}],  the largest contribution below the spectral gap is due to the absorption of phase modes, with an
admixture of amplitude and charge excitations. Thus, one would expect that exact shape and strength of the extra absorption could be affected by the inclusion of long-range Coulomb forces. Nonetheless, the main trends of the experiments are already captured by our calculations, see Fig.\ {\ref{fig2}, suggesting that screening effects are probably not so dramatic in real systems. If this is just a consequence of the low value of the plasma mode, due to the low superfluid density, or it is an inherent effect of the local character of the screened phase modes at strong disorder is an interesting open question that we leave for future investigation.

  \acknowledgments{We acknowledge useful discussions with P. Armitage,
    and we thank the authors of Ref.\ [\onlinecite{armitage15}] for providing us with the experimental data shown in the figures. This work has been partly supported by the Italian
MIUR under projects FIRB-HybridNanoDev-RBFR1236VV,
PRINRIDEIRON-2012X3YFZ2,  and  Premiali-2012  AB-
NANOTECH. }

\appendix
\section{Mattis-Bardeen formula for frequency dependent DOS}\label{sec:a1}
The formula Eq. (\ref{eq:iw}) for the regular part $\sigma_1(\omega)$ given in the main text
is valid for a symmetric DOS $N(\omega)$ around the chemical potential. The
general formula reads
\begin{widetext}
\begin{eqnarray}
  I(\omega)&=& \frac{1}{2\omega} \int_\Delta^\infty\!dE\,\frac{f(E)-f(E+\omega)}
  {\epsilon_1\epsilon_2} \left\lbrace \left\lbrack E(E+\omega) 
  +\epsilon_1\epsilon_2 +\Delta^2\right\rbrack \left\lbrack N(\epsilon_1)N(\epsilon_2)
  +N(-\epsilon_1)N(-\epsilon_2)\right\rbrack \right. \nonumber\\
  &+& \left. \left\lbrack E(E+\omega) -\epsilon_1\epsilon_2 +\Delta^2\right\rbrack 
  \left\lbrack N(\epsilon_1)N(-\epsilon_2)
  +N((-\epsilon_1)N(\epsilon_2)\right\rbrack\right\rbrace \nonumber\\
  &-& \frac{1}{4\omega}\int_{\Delta-\omega}^{-\Delta}\!dE\,\frac{1-2f(E+\omega)}
  {\varepsilon_1\varepsilon_2}\left\lbrace \left\lbrack E(E+\omega) 
  +\epsilon_1\epsilon_2 +\Delta^2\right\rbrack \left\lbrack N(\epsilon_1)N(\epsilon_2)
  +N((-\epsilon_1)N(-\epsilon_2)\right\rbrack \right. \nonumber\\
  &+& \left. \left\lbrack E(E+\omega) -\epsilon_1\epsilon_2 
  +\Delta^2\right\rbrack \left\lbrack N(\epsilon_1)N(-\epsilon_2)
  +N(-\epsilon_1)N(\epsilon_2)\right\rbrack\right\rbrace \label{eq:amb}\,.
\end{eqnarray}
\end{widetext}

As one can immediately see, when $N(\omega)=N_0=const$ one recovers back the standard MB formula. In particular in the limit $T\rightarrow 0$ one recovers from Eq.\ (\ref{eq:gw}) that $\Gamma(\omega)=N_0^2$. By computing also $I(\omega)$ one finds:
\begin{widetext}
\begin{eqnarray}
\label{sigma1mb}
\frac{\sigma_{1s}}{\sigma_{1n}}&=& \frac{2}{\omega} \int_\Delta^\infty\!dE\,\left[f(E)-f(E+\omega)\right]
\frac{\left\lbrack E(E+\omega) +\Delta^2\right\rbrack }{ \sqrt{E^2-\Delta^2}\sqrt{(E+\omega)^2-\Delta^2}}\nonumber\\
&-&\frac{\Theta(\omega-2\Delta)}{\omega}\int_{\Delta-\omega}^{-\Delta}\!dE\,
\left[1-2f(E+\omega)\right]
\frac{\left\lbrack E(E+\omega) +\Delta^2\right\rbrack}{ \sqrt{E^2-\Delta^2}\sqrt{(E+\omega)^2-\Delta^2}}\\\
\label{sigma2mb}
\frac{\sigma_{2s}}{\sigma_{1n}}&=& \frac{2}{\omega} \int_{-\Delta,\Delta-\omega}^\infty\!dE\,[1-2f(E+\omega)]
\frac{\left\lbrack E(E+\omega) +\Delta^2\right\rbrack}{\sqrt{\Delta^2-E^2}\sqrt{(E+\omega)^2-\Delta^2}} 
\end{eqnarray}
\end{widetext}
where together with the real part we also wrote explicitly the imaginary part of the conductivity for the standard MB case. In the expression for $\sigma_2$ the lower limit of integration is taken as the largest between the two values $-\Delta$ and $\Delta-\omega$.

\section{Analysis of $T_c=9.2 K$ data}\label{sec:apb}
Fig. \ref{fig5} shows the fits from extended Mattis-Bardeen theory for a
disordered $T_c=9.2 K$ sample. The analysis of the data is given in
Sec. \ref{sec:mb} of the main text.
\begin{figure}[htb]
\includegraphics[width=8.5cm,clip=true]{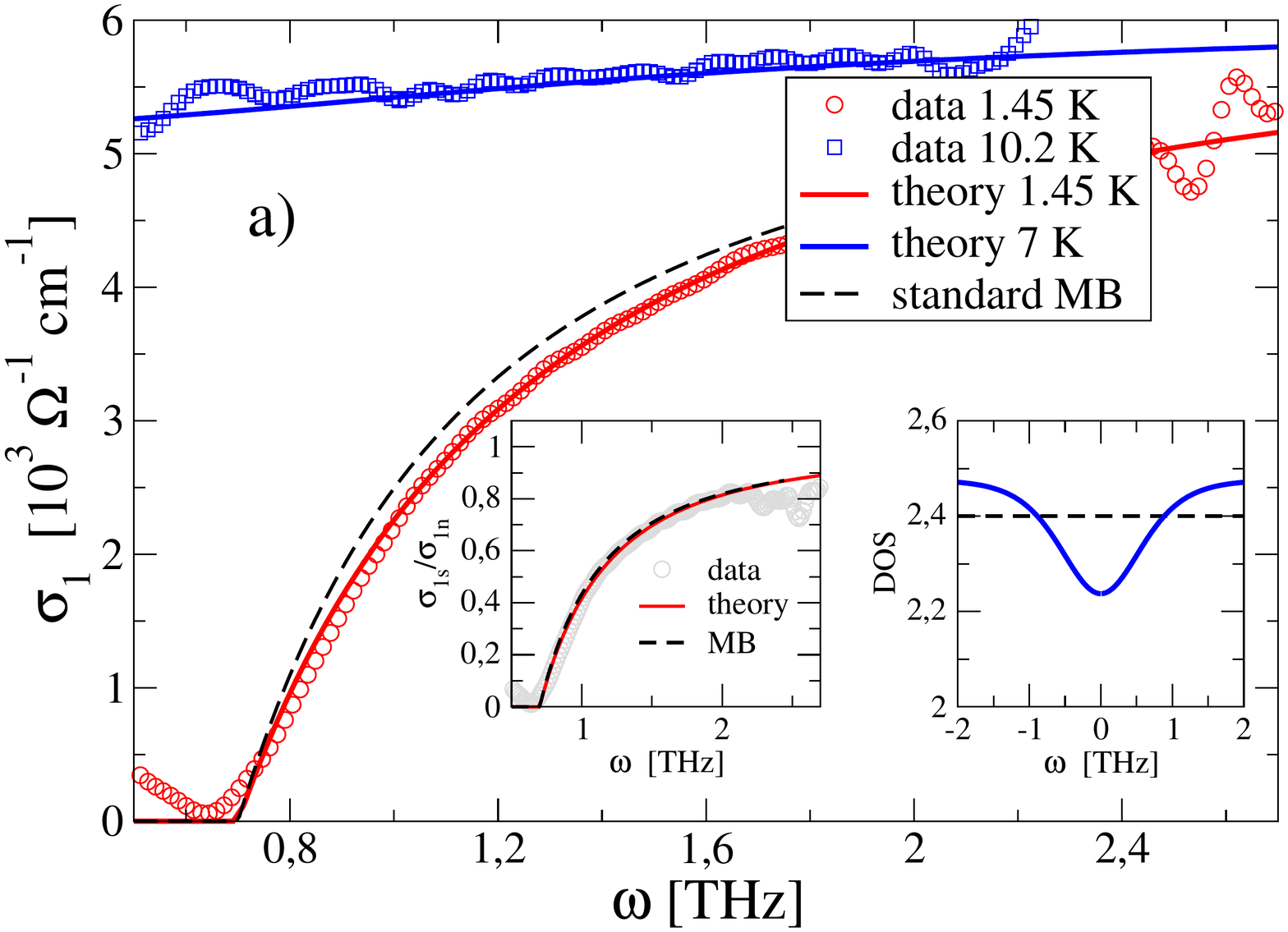}
\includegraphics[width=8.5cm,clip=true]{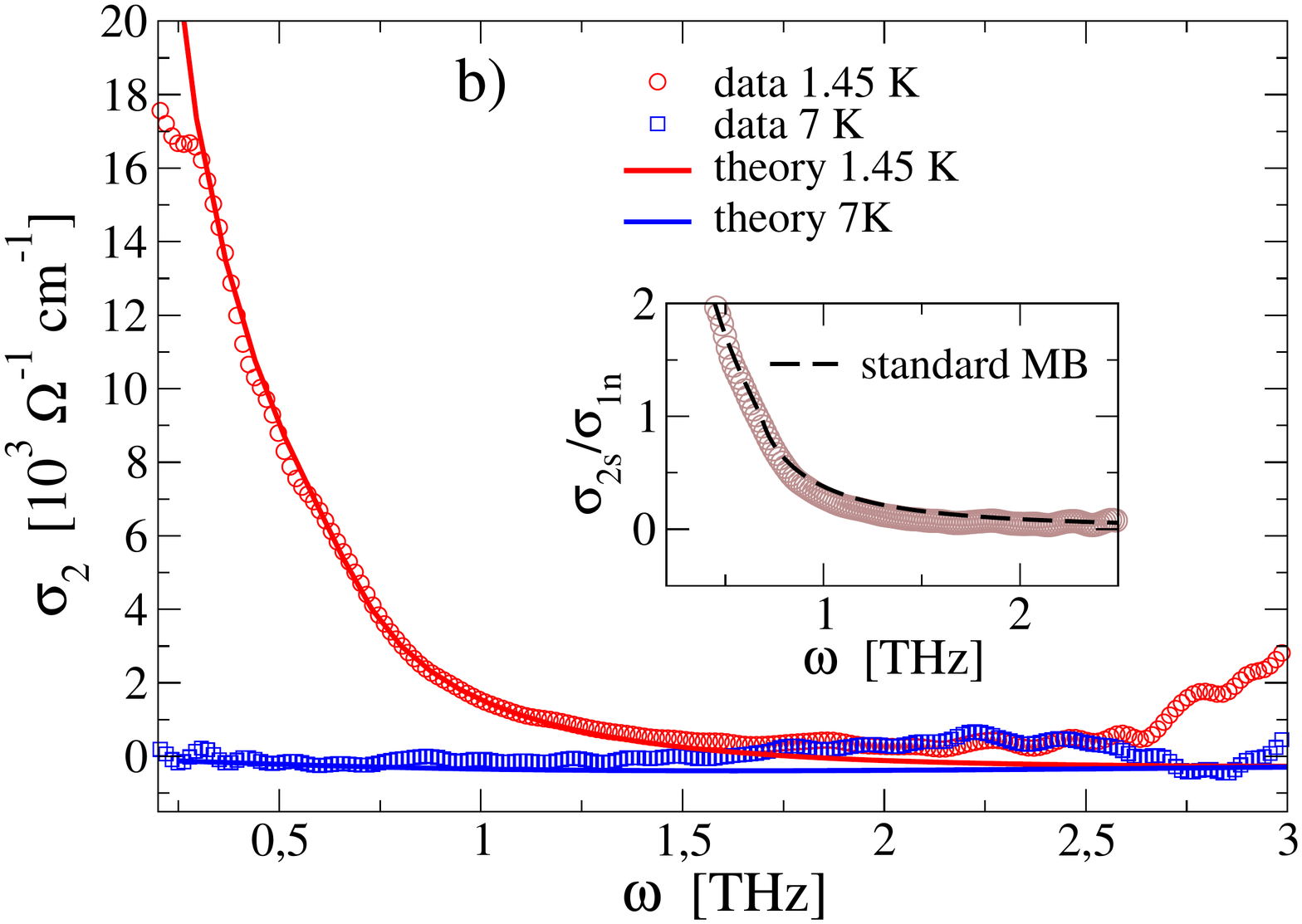}
\caption{Same as Fig. \ref{fig4} but for  
a NbN sample with $T_c=9.2 K$. Parameters for the DOS Eq. \ref{eq:dos}:
$\alpha=2.24$, $\beta=0.24$, $\Omega=0.75 THz$. Superconducting gap parameter:
$\Delta=0.35 THz$. 
Experimental data by courtesy of the authors of Ref.\ [\onlinecite{armitage15}].}
\label{fig5}
\end{figure}

\end{document}